# "Perfectly" curvilinear profiles for binding as determined by ITC may in fact be multiphasic

Per Nissen

Norwegian University of Life Sciences
Department of Ecology and Natural Resource Management
P. O. Box 5003, NO-1432 Ås, Norway

per.nissen@nmbu.no

2016

# Abstract

In a structural analysis of the proteasome activator PafE in *Mycobacterium tuberculosis*, the binding of the activator or shorter constructs to the 20S proteasome core particle (20S CP) or derivatives was measured by isothermal titration calorimetry (Bai et al. *Proc Natl Acad Sci USA* 113: E1983-E1992. 2016). The data were fitted by the authors by nonlinear least squares to give curvilinear profiles that, at least in part, appear to fit the data very well. However, reanalysis of the data shows that the profiles are much better (P < 0.001) represented as multiphasic, i.e. by a series of straight lines separated by discontinuous transitions, often in the form jumps, than by the conventional curvilinear profiles.

# Introduction

In addition to multiphasic profiles for ion uptake in plants (Nissen 1971, 1974, 1991, 1996), such profiles have been recently (Nissen 2015a,b, 2016a,b) reported for many other processes and phenomena. In the present paper, data (Bai et al. 2016) for binding as determined by ITC will be reanalyzed to statistically compare the fits to curvilinear profiles with the fits to multiphasic profiles.

Original data were kindly provided by Huilin Li. In addition to the r values, slopes ± SE (or only slopes) are given on the plots.

# Reanalysis

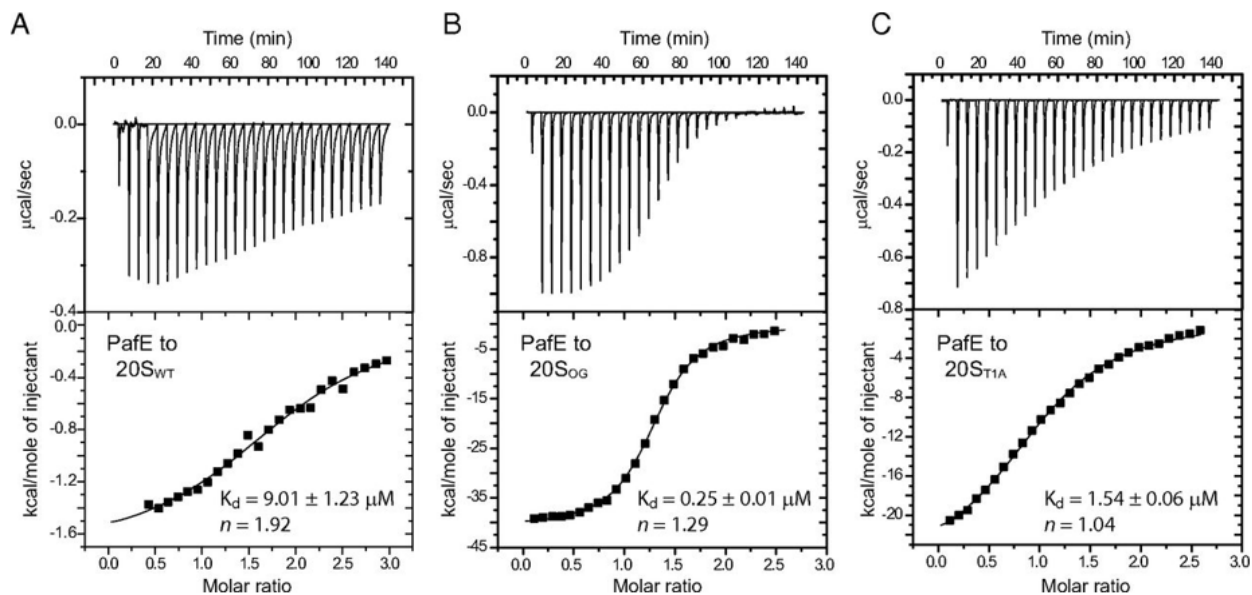

**Fig. 1.** Fig. 3 in Bai et al. (2016). In vitro binding of PafE to *Mtb* 20S CPs as measured by isothermal titration calorimetry. *Lower* panels: (A) 480 μM PafE into 33 μM *Mtb* 20S$_{WT}$, and (B and C) 85 μM PafE into 6.7 μM *Mtb* 20S$_{OG}$ and 20S$_{T1A,}$ respectively. See also original legend.

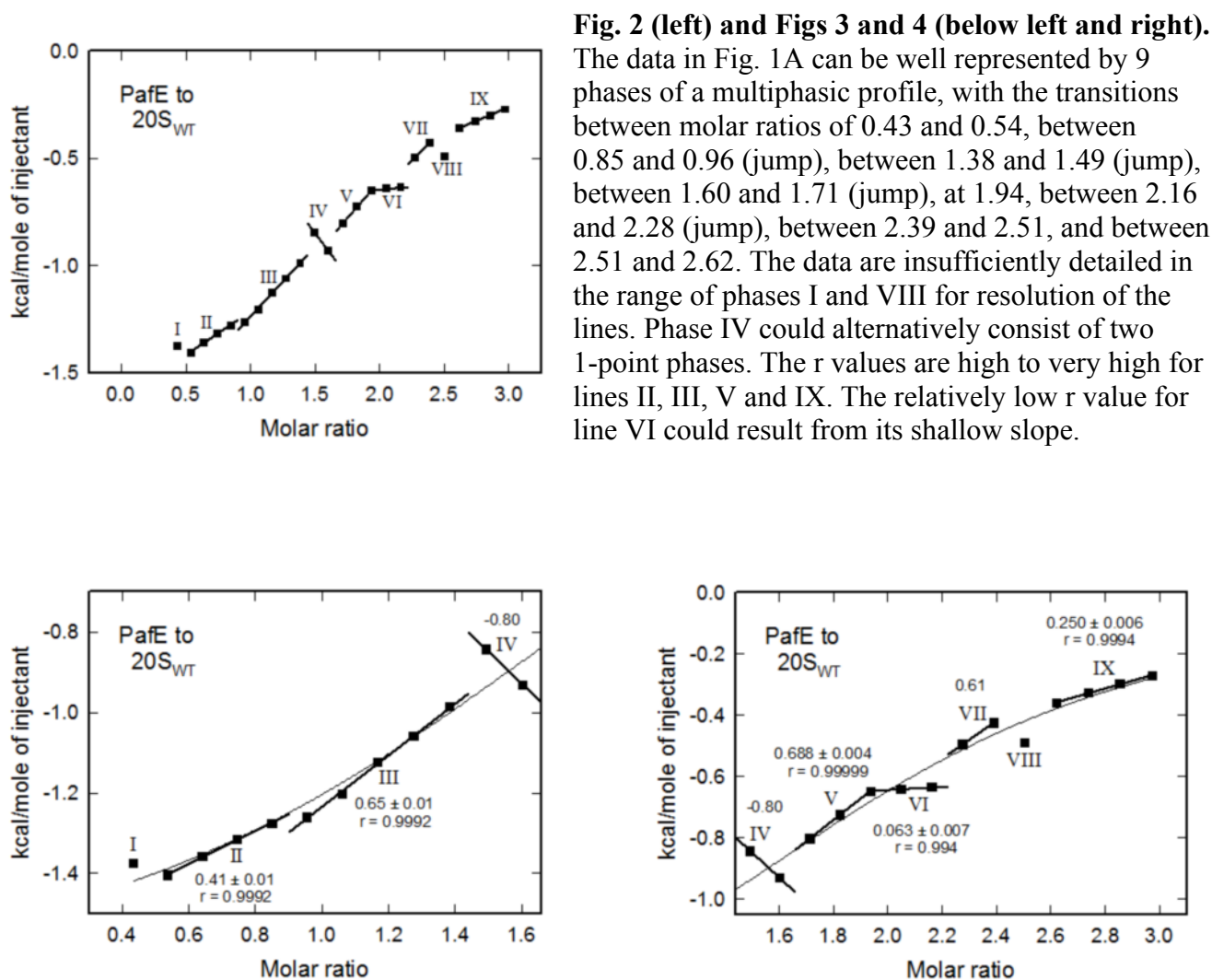

**Fig. 2 (left) and Figs 3 and 4 (below left and right).** The data in Fig. 1A can be well represented by 9 phases of a multiphasic profile, with the transitions between molar ratios of 0.43 and 0.54, between 0.85 and 0.96 (jump), between 1.38 and 1.49 (jump), between 1.60 and 1.71 (jump), at 1.94, between 2.16 and 2.28 (jump), between 2.39 and 2.51, and between 2.51 and 2.62. The data are insufficiently detailed in the range of phases I and VIII for resolution of the lines. Phase IV could alternatively consist of two 1-point phases. The r values are high to very high for lines II, III, V and IX. The relatively low r value for line VI could result from its shallow slope.

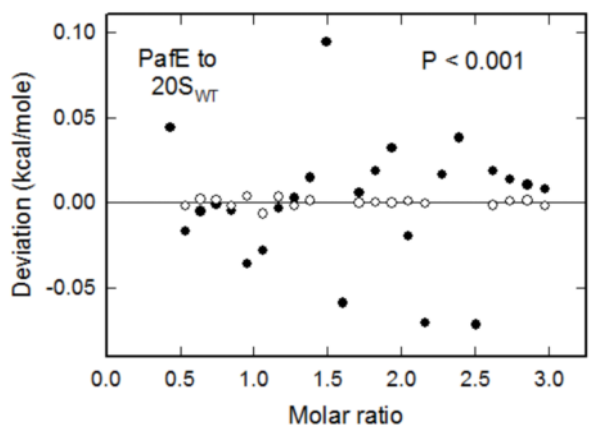

**Fig. 5.** Plot of deviates for the data in Fig. 1A. The deviations (filled circles) from the curvilinear profile are on the whole much larger than the deviations (open circles) from the multiphasic profile. The difference is highly significant (by the Mann-Whitney rank sum test).

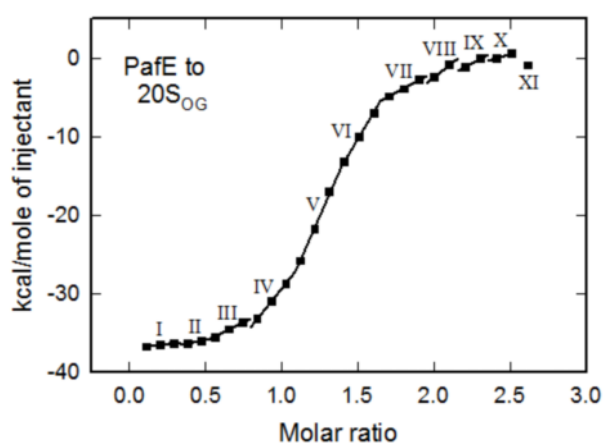

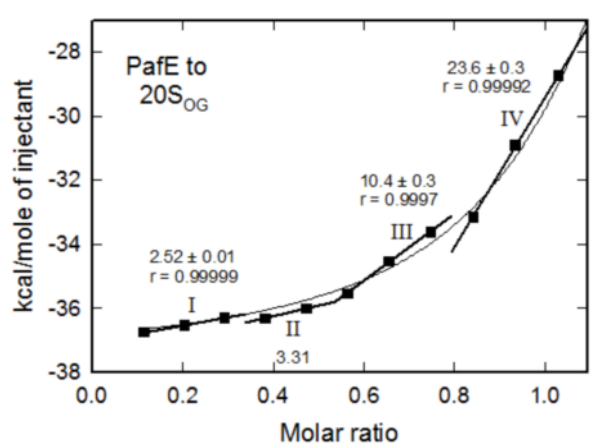

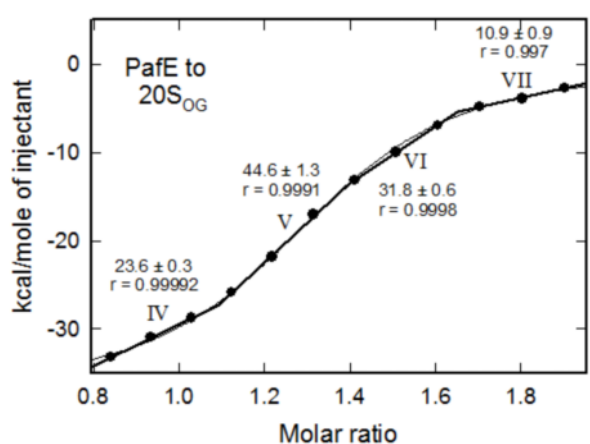

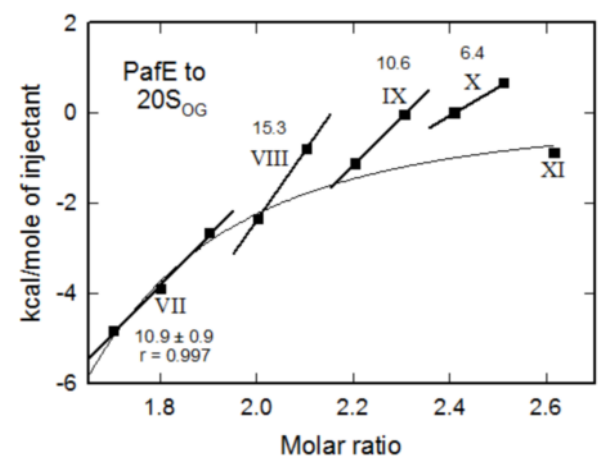

**Figs 6** and **7 (**upper left and right) and **Figs 8** and **9** (lower left and right). The data in Fig. 1B can be well represented by 11 phases, with transitions between molar ratios of 0.29 and 0.38 (jump), at 0.54, between 0.75 and 0.84 (jump), at 1.09, 1.41 and 1.65, between 1.90 and 2.00 (jump), between 2.10 and 2.21 (jump), between 2.31 and 2.41 (jump), and between 2.51 and 2.62. Very high r values for lines I, III, IV, V and VI, lower r value for line VII. The point for phase XI is not shown in Fig. 1B, but is on the printout from the calorimeter.

..

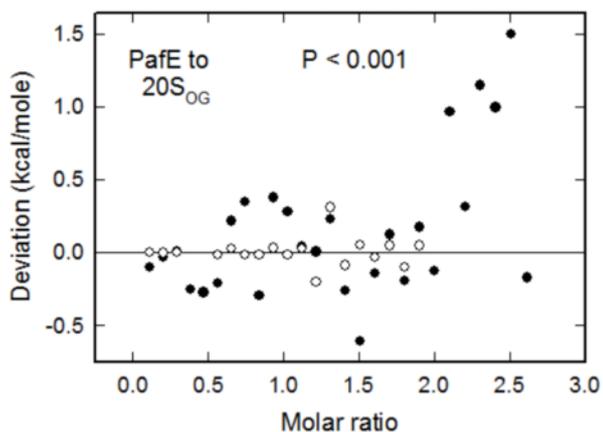

**Fig. 10.** Plot of deviates for the data in Fig. 1B. The deviations from the curvilinear profile are on the whole much larger than the deviations from the multiphasic profile. The difference is highly significant.

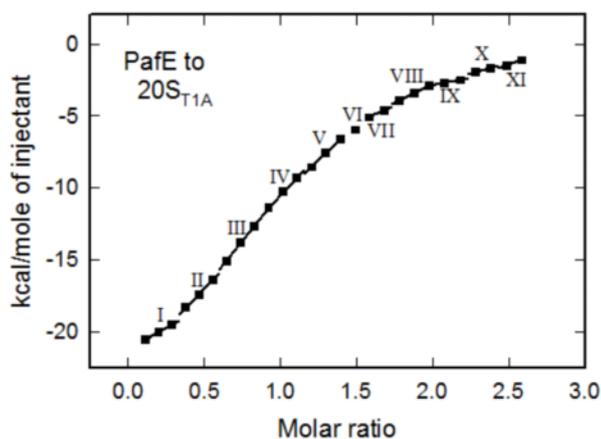

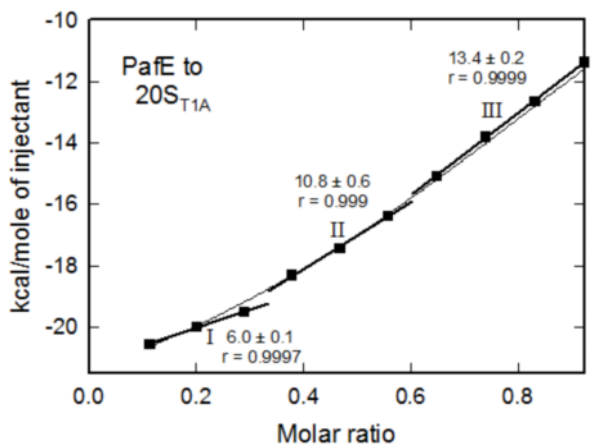

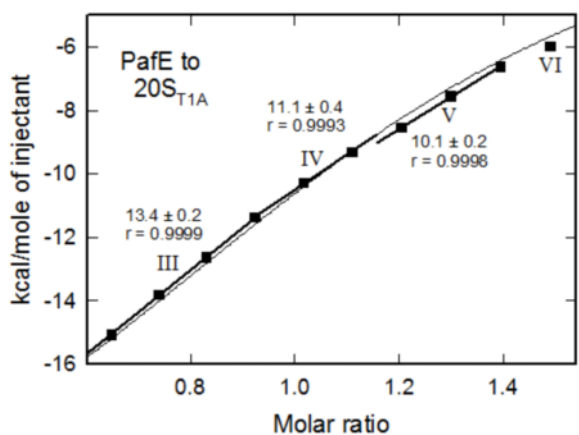

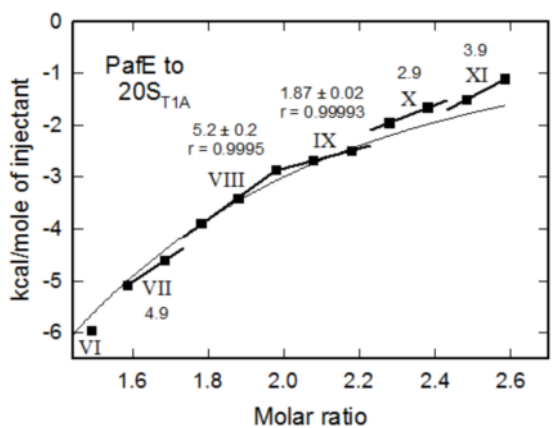

**Figs 11** and **12** (upper left and right) and **Figs 13** and **14** (lower left and right). As also for the data in Fig. 1B, the data in Fig. 1C can be well represented by 11 phases, with the transitions between molar ratios of 0.29 and 0.38 (jump), between 0.56 and 0.65 (jump), at 0.92, between 1.11 and 1.20 (jump), between 1.39 and 1.49, between 1.49 and 1.59, between 1.68 and 1.78 (jump), at 1.98, between 2.18 and 2.28 (jump), and between 2.38 and 2.48 (jump). The data are insufficiently detailed in the range of phase VI for resolution of the line. The r values are high to exceedingly high for all lines with three or more points. Lines VII and VIII are parallel, as are, approximately, lines X and XI. The non-adjacent lines II and IV are also parallel.

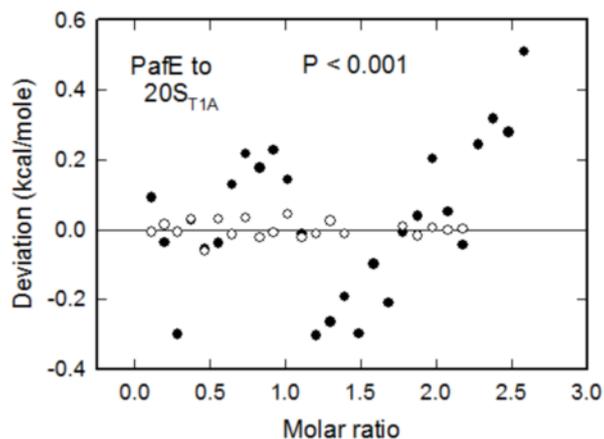

**Fig. 15.** Plot of deviates for the data in Fig. 1C. As also for the data in Figs 1A and 1B, the deviations from the curvilinear profile are on the whole much larger than the deviations from the multiphasic profile. The difference is highly significant.

## Conclusion

Profiles for binding are often determined by isothermal titration calorimetry and represented as curvilinear. The fits to curvilinear profiles may quite often appear to be very good, if not perfect (cf. Figs 1B and 1C). However, the present reanalysis reveals that the data in Fig. 1 are much better ($P < 0.001$) represented as multiphasic than as curvilinear.

The finding that profiles for binding are better represented by multiphasic profiles than by curvilinear profiles agrees with previous findings (Nissen 2015a), both for binding as determined by ITC (data from Lunelli et al. 2009) and as determined by anisotropy (data from Burge et al. 2009).

**Acknowledgment** – I am very grateful to Bob Eisenberg for his continued interest and encouragement.